\documentclass{article}

\usepackage{arxiv}

\usepackage[utf8]{inputenc} 
\usepackage[T1]{fontenc}    
\usepackage{hyperref}       
\usepackage{url}            
\usepackage{booktabs}       
\usepackage{amsfonts}       
\usepackage{nicefrac}       
\usepackage{microtype}      
\usepackage{lipsum}
\usepackage{graphicx}
\graphicspath{ {./images/} }
\usepackage{xcolor}
\usepackage{blindtext}
\usepackage{tabularx}
\usepackage{multicol}
\usepackage{balance}
\usepackage{subcaption}
\usepackage{multirow}
\usepackage{balance}
\usepackage{array}
\newcolumntype{C}[1]{>{\centering\arraybackslash}p{#1}}

\usepackage{pifont}
\usepackage{listings}

\newif\ifarxiv
\arxivtrue  

\definecolor{codegreen}{rgb}{0,0.6,0}
\definecolor{codegray}{rgb}{0.5,0.5,0.5}
\definecolor{codepurple}{rgb}{0.58,0,0.82}
\definecolor{backcolour}{rgb}{0.95,0.95,0.92}

\lstdefinestyle{mystyle}{
    backgroundcolor=\color{backcolour},   
    commentstyle=\color{codegreen},
    keywordstyle=\color{magenta},
    numberstyle=\tiny\color{codegray},
    stringstyle=\color{codepurple},
    basicstyle=\ttfamily\footnotesize,
    breakatwhitespace=false,         
    breaklines=true,                 
    captionpos=b,                    
    keepspaces=true,                 
    numbers=left,                    
    numbersep=5pt,                  
    showspaces=false,                
    showstringspaces=false,
    showtabs=false,                  
    tabsize=2
}

\title{Evaluating Kubernetes Performance for GenAI Inference:  From Automatic Speech Recognition to LLM Summarization}

\author{
Sai Sindhur Malleni\thanks{This work was performed while the author was at Red Hat. The author is now with NVIDIA.} \\
Red Hat \\
Boston, MA, US \\
\texttt{smalleni@redhat.com}
\And
Raúl Sevilla \\
Red Hat \\
Madrid, Spain\\
\texttt{rsevilla@redhat.com}
\And
Aleksei Vasilevskii \\
Red Hat \\
Munich, Germany \\
\texttt{avasilev@redhat.com}
\And
José Castillo Lema \\
Red Hat \\
Madrid, Spain\\
\texttt{jlema@redhat.com} \\
\And
André Bauer \\
Illinois Institute of Technology \\
Chicago, IL, US \\
\texttt{andre.bauer@iit.edu}
}

\begin{document}
\maketitle
\begin{abstract}
As Generative AI (GenAI), particularly inference, rapidly emerges as a dominant workload category, the Kubernetes ecosystem is proactively evolving to natively support its unique demands. This industry paper demonstrates how emerging Kubernetes-native projects can be combined to deliver the benefits of container orchestration, such as scalability and resource efficiency, to complex AI workflows. We implement and evaluate an illustrative, multi-stage use case consisting of automatic speech recognition and summarization. First, we address batch inference by using Kueue to manage jobs that transcribe audio files with Whisper models and Dynamic Accelerator Slicer (DAS) to increase parallel job execution. Second, we address a discrete online inference scenario by feeding the transcripts to a Large Language Model for summarization hosted using llm-d, a novel solution utilizing the recent developments around the Kubernetes Gateway API Inference Extension (GAIE) for optimized routing of inference requests. Our findings illustrate that these complementary components (Kueue, DAS, and GAIE) form a cohesive, high-performance platform, proving Kubernetes' capability to serve as a unified foundation for demanding GenAI workloads: Kueue reduced total makespan by up to 15\%; DAS shortened mean job completion time by 36\%; and GAIE working in conjunction with llm-d improved tail Time to First Token latency by up to 90\% even under high loads.
\end{abstract}


\section{Introduction}

The global shift toward Artificial Intelligence (AI) has fundamentally changed the landscape of cloud computing. Modern AI workloads are intricate, spanning complex training pipelines that build sophisticated models and high-throughput real-time inference serving that deploys them. While the training phase requires substantial, often bursty, computational resources, the industry's focus has increasingly shifted toward deployment. This is especially true for Generative AI (GenAI), which was exhibiting an 18.7\% increase in private investment from 2023 alone, according to the Stanford HAI AI Index Report~\cite{AIIndex2025}. Critically, the primary infrastructure challenge today lies in serving this massive demand; as noted by ACM Queue, ``AI: It's All About Inference Now''~\cite{Gschwind}. Inference workloads require dynamic access to heterogeneous and often expensive resources, particularly GPUs and specialized accelerators~\cite{grandview2025_cloudAI}. To successfully manage the entire lifecycle of these multi-stage AI applications from provisioning data to model deployment, a sophisticated and extensible orchestration layer abstracting the underlying hardware and providing declarative management is required.

This is where orchestration systems like Kubernetes become essential. Initially rising to prominence through the microservices revolution, Kubernetes has established itself as the de facto standard for container orchestration, automating the deployment, scaling, and management of containerized applications~\cite{burns2016borg}. According to the Cloud Native Computing Foundation’s most recent annual survey, 91\% of organizations use containers for production~\cite{cncfannualsurvey}. Given its ability to efficiently scale to meet varying computational demands and its rich collection of AI frameworks and tools, Kubernetes also has emerged as a key enabler for customers to run large scale AI. However, as AI models continue to evolve in size and complexity, they require advanced capabilities beyond Kubernetes’ traditional boundaries. Traditional scheduling and resource management APIs were not designed to handle fine-grained GPU resource slicing, complex scheduling requirements for distributed training, or the specific traffic patterns of model serving endpoints.  

This industry paper demonstrates how emerging Kubernetes-native pro\-jects can be combined to deliver the full benefits of container orchestration, such as scalability and resource efficiency, to complex AI inference workflows. We argue that three complementary projects explicitly address the distinct needs of such workloads by focusing on key stages of the deployment and serving pipeline: Kueue~\cite{kueue} for efficient batch scheduling, Dynamic Accelerator Slicer~\cite{das} (DAS) for resource optimization, and the newly emerged Gateway API Inference Extension~\cite{kubernetes-sigs-gateway-api-inference-extension} (GAIE) for distributed inference performance on Kubernetes.

To showcase their synergy, we implement and evaluate an illustrative, multi-stage use case processing a dataset of corporate earnings calls~\cite{earnings22}, a workflow that reflects common industry usage patterns. First, we address batch inference by using Kueue to manage jobs that transcribe audio files with Whisper models, a core component of Automatic Speech Recognition (ASR). ASR has emerged as a critical enabling technology, with the ASR software market valued at USD 5.49 billion in 2024 \cite{mrfuture2024asr}. We demonstrate significant efficiency gains for these workloads by integrating DAS, which leverages NVIDIA MIG (Multi-Instance GPU) slicing to dramatically increase parallel job execution on the same hardware. Second, we address a discrete online inference scenario by feeding the generated transcripts to a Large Language Model (LLM) for summarization. This service is hosted using llm-d~\cite{llmd}, a novel solution utilizing the recent developments around the GAIE for optimized routing of inference requests. We showcase responsiveness gains by applying its vLLM-Optimized Inference Scheduler, which employs a ``Precise Prefix-Cache Aware Scheduling'' strategy.

Our findings illustrate that these complementary components (Kueue, DAS, and GAIE) complement core Kubernetes to form a cohesive, high-performance platform, proving Kubernetes' capability to serve as a unified foundation for demanding GenAI inference workloads. Specifically, Kueue reduced the total makespan by up to 15\%, DAS decreased the mean job completion time by 36\%, and GAIE lowered inference latency by as much as 6 seconds while improving the time to first token by more than sixfold. It is important to note that while model accuracy is a crucial metric from an AI perspective, the scope of this industry paper is focused exclusively on the system-level performance, efficiency, and operational overhead of the underlying orchestration infrastructure.

The remainder of this industry paper is structured as follows:
Section~\ref{sec:back} introduces Kubernetes primitives for AI workloads and related
work. Section~\ref{sec:approach} outlines the measurement setup, deployed inference AI solutions, and performed test cases. Section~\ref{sec:eva}
investigates the results of the analyzed solutions.
Finally, Section~\ref{sec:conclusion} summarizes the paper and Section~\ref{sec:future} discusses future work directions.



\section{Background}\label{sec:back}

This chapter establishes the necessary background on AI workloads and Kubernetes and discusses related work.

\subsection{Kubernetes Primitives for AI Workloads}

Modern AI training and inference pipelines build on several Kubernetes primitives that expose accelerators, respect hardware topology and NUMA affinities, and enforce resource limits. The \textit{device plugin} framework allows vendors to advertise GPUs, FPGAs, DPUs and other accelerators to the scheduler so that pods can request these resources in a portable way~\cite{k8sDevicePlugin}.
The \textit{Topology Manager} coordinates CPU and device placement and supports policies up to \textit{single-NUMA-node} to improve locality and latency for latency-sensitive or accelerator-heavy pods~\cite{k8s_topology_manager}. \textit{Dynamic Resource Allocation (DRA)} generalizes persistent‑volume claims to generic devices (e.g., GPUs or DPUs), introducing device classes and claims so operators can safely share accelerators; production support in cloud distributions is emerging~\cite{k8s_dra}. Finally, \textit{resource quotas}, priority classes and related limit mechanisms remain essential for fair sharing and multi‑tenancy in AI clusters~\cite{k8s_quotas}.

On the device side, NVIDIA MIG partitions a single GPU into multiple isolated slices with dedicated memory and compute engines, allowing heterogeneous and multi‑tenant inference services to share an expensive accelerator efficiently~\cite{nvidia_mig}. When strict isolation is unnecessary, \textit{GPU time‑slicing} multiplexes workloads over the entire device; both modes are available through the NVIDIA GPU Operator and device‑plugin stacks~\cite{nvidia_gpu_operator}. These primitives underpin the singular inference solutions evaluated in this paper.

\subsection{AI Workload Orchestration on Kubernetes}

Integral to orchestrating AI workloads on Kubernetes are three capabilities: job queuing and scheduling, accelerator slicing, and multi-node inference orchestration for distributed inference.

\subsubsection{Scheduling (Queueing and Admission).} The default Kubernetes scheduler is insufficient for complex AI workloads, as it lacks job-level admission control and gang scheduling. To address this, emerging systems extend the platform. These range from comprehensive commercial platforms like NVIDIA Run.ai, which provides a full GPU orchestration layer with queueing, preemption, and hierarchical quota management~\cite{runai}, to open-source, community-driven schedulers.

\subsubsection{Accelerator Slicing.} Accelerator sharing requires both a low-level partitioning technology and a high-level orchestration API. The primary partitioning methods are NVIDIA MIG, for hardware-isolated slices, and time-slicing, for software-based sharing~\cite{nvidia_mig}. The traditional Kubernetes device plugin model, however, only supports static, pre-configured slices. To address this, operator-based solutions like the \textit{Dynamic Accelerator Slicer} (DAS) were developed to automate ``just-in-time'' slicing. DAS dynamically creates and destroys MIG partitions to match workload demands, improving utilization~\cite{das}. This provides a bridge to the official, upstream API, DRA, which is maturing to provide a standardized, abstract framework for requesting and sharing such slices natively in Kubernetes~\cite{k8s_dra}.

\subsubsection{Distributed Inference.} Foundation models increasingly exceed a single node's memory and accelerator budget, pushing inference toward distributed execution. The \textit{Gateway API Inference Extension} (GAIE) provides a new layer of intelligence to Kubernetes networking, tailored for generative AI workloads~\cite{kubernetes-sigs-gateway-api-inference-extension}. Built on the standard Gateway API, it introduces domain-specific constructs to dynamically route requests based on live model and accelerator metrics (e.g., queue length, prefix cache hits, Low-Rank Adaptation (LoRA)~\cite{hu2022lora} adapter availability). This model-aware routing enables higher GPU utilization and lower tail latency.

This paper focuses on the optimizations and performance gains provided by this extension and its inference schedulers. To conduct this analysis, we utilize \textit{vLLM}~\cite{kwon2023vllm} as the high-performance model server backend. While our distributed inference solution is deployed using the \textit{llm-d} framework~\cite{llmd}, this paper's evaluation is focused specifically on the optimizations provided by the  GAIE itself, which is detailed in Section \ref{sec:igw}.






\subsection{Related Work}

The rise of Kubernetes as the foundational platform for managing AI workloads has spurred extensive research focused on enhancing its capabilities in scheduling, resource management, and model serving.

\subsubsection{Scheduling and Orchestration}

Research on scheduling in Kubernetes aims to optimize performance and resource utilization for AI/ML tasks. Surveys by Rejiba et al.~\cite{schedulingsurvey} and Ye et al.~\cite{DLschedulingsurvey} detail common challenges and solutions, with the latter focusing specifically on deep learning training and inference. To improve scheduling effectiveness, various learning-based frameworks have been proposed: PAX~\cite{PAX} introduces a machine learning-driven scheduler for performance and energy efficiency, and KaiS~\cite{KaiS} applies a similar approach in edge-cloud environments to boost long-term performance. For distributed deep learning, systems like ElasticDL~\cite{ElasticDL} automate deployment and elastic scaling, while MuxFlow~\cite{Muxflow} refines allocation via dynamic multiprocessor allocation and matching-based mechanisms.

\subsubsection{Resource Slicing and Sharing}

Another line of work addresses the efficient slicing and sharing of hardware accelerators, particularly GPUs. Tools like nvshare~\cite{nvshare} enable practical GPU sharing without memory-size constraints, and DISC~\cite{Disc} dynamically adjusts GPU time slices for hyperparameter tuning jobs. For more advanced management, Jormungandr~\cite{Jormungandr} optimizes the allocation of NVIDIA's MIG based on user-defined utility functions, and KIS-S~\cite{KISS} provides a GPU-aware autoscaler tailored for inference workloads. Beyond GPUs, work also extends to container partitioning for distributed ML~\cite{2DFS} and integrating other accelerators, such as extending KServe for FPGA-based workloads~\cite{KServeFPGA}.

\subsubsection{Distributed Inference}
Distributed inference has emerged as a pivotal approach for deploying large-scale models across multiple nodes. Several systems address the challenges of partitioning and coordinating inference workloads. Alpa~\cite{alpa} introduces compiler-based automated model parallelism for distributed serving. PetS~\cite{zhou2022pets} and DistServe~\cite{zhong2024distserve} specifically target disaggregated prefill and decode stages in LLM inference to enhance resource utilization and throughput. Xu et al.~\cite{LLMinferenceServing} comprehensively investigate how Kubernetes can optimize LLM serving performance. ServerlessLLM~\cite{serverlessllm} proposes elastic scaling for distributed LLM workloads. Comparative evaluations~\cite{modelservingcomparison} assess orchestration frameworks like KServe for multi-node deployments.

\subsubsection{Delimitation}
Prior work has largely focused on novel scheduling algorithms or hardware specific resource sharing techniques. In contrast, relatively few studies present an end to end analysis of the operational overheads, performance, and scaling behavior of a full stack, state of the art AI workload running on Kubernetes. This work addresses that gap by evaluating a complete inference pipeline rather than isolated system components or microbenchmarks.

We build and evaluate an end to end pipeline consisting of automatic speech recognition followed by text summarization. For the ASR stage, we use OpenAI’s Whisper~\cite{radford2023robust}, which is available in multiple model sizes and serves as a representative real world ASR workload. ASR systems are increasingly important in modern computing, supporting applications such as voice assistants, accessibility tools, and real time transcription. These workloads are both computationally intensive and latency sensitive, making them well suited for studying system level performance and scalability~\cite{prabhavalkar2024endtoend,park2022low}.

The pipeline is evaluated under two different inference regimes. First, we study batch inference, where Kueue is used to manage large volumes of transcription jobs and Dynamic Accelerator Slicer (DAS) is used to improve GPU utilization and increase parallel job execution. Second, we examine a real-time inference scenario in which the generated transcripts are passed to a large language model for summarization. This stage is deployed using llm-d, a Kubernetes native inference serving solution that builds on recent developments in the Kubernetes Gateway API Inference Extension (GAIE) to enable efficient request routing and scalable online inference.

This work deliberately does not propose new scheduling algorithms or model architectures. Instead, it focuses on characterizing the system level behavior of modern Kubernetes native components when running realistic AI workloads, and on highlighting the performance, scalability, and operational trade offs that arise in production like environments.

\section{Measurement Setup}\label{sec:approach}

\begin{figure}[htb!]
    \centering
    \ifarxiv
      \includegraphics[width=0.75\linewidth]{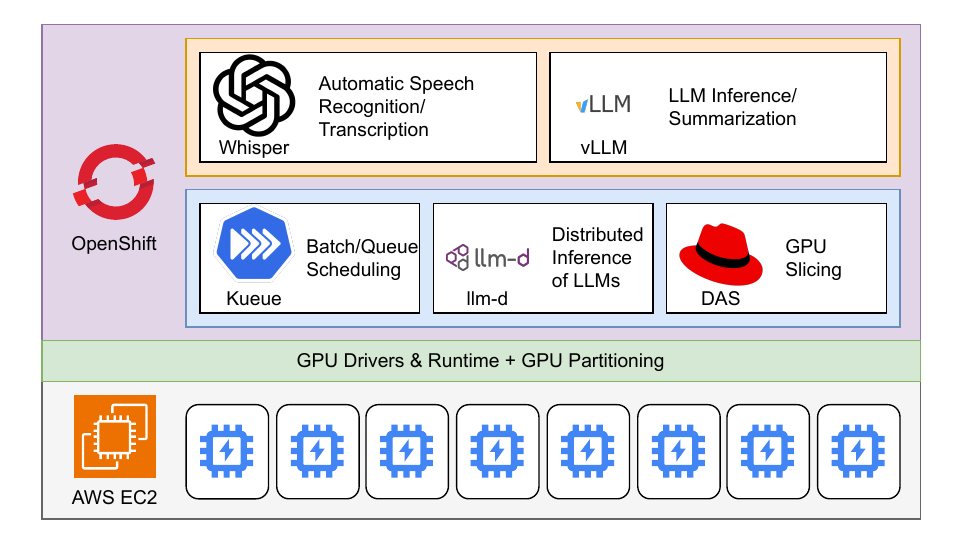}
    \else
    \includegraphics[width=\linewidth]{figures/architecture.pdf}
    \fi
    \caption{Overview of the technology stack.}
    \label{fig:architecture}
\end{figure}

The complete utilized technology stack is depicted in Figure~\ref{fig:architecture} and described in the following sections.

\subsection{Infrastructure Stack}\label{sec:stack}

To demonstrate how emerging Kubernetes-native projects can be combined to maximize the benefits of container orchestration, we present an illustrative, multi-stage use case that processes a dataset of corporate earnings calls~\cite{earnings22}. In this section, we elaborate on the solutions employed for scheduling, accelerator slicing, and multi-node inference. Table~\ref{tab:ai_k8s_comparison} offers a concise overview of each solution.

\subsubsection{Scheduling (Queueing \& Admission)}

\textit{Kueue}~\cite{kueue} is an open-source project under the Kubernetes Scheduling Special Interest Group that extends native Kubernetes scheduling to support advanced batch and job management. Unlike the standard Kubernetes Job controller, which treats each job independently, \textit{Kueue} introduces a hierarchical queuing model composed of \emph{LocalQueues} and \emph{ClusterQueues}. This abstraction enables cluster operators to declaratively define resource boundaries, priority classes, and admission controls for batch workloads, thereby improving fairness and utilization across shared clusters.

A \textit{LocalQueue} is namespace-scoped and serves as the entry point for job submissions within that namespace. Each \textit{LocalQueue} is bound to one or more \textit{ClusterQueues}, which operate at the cluster scope and represent the global scheduling domains responsible for admitting jobs based on cluster-wide policies and resource availability. This decoupling between job submission via \textit{LocalQueues} and resource management via \textit{ClusterQueues} allows individual teams to operate independently while maintaining centralized control over compute resources.

\begin{table*}[ht]
\centering
\caption{Compact comparison of Kueue, DAS, and llm-d for AI workloads}
\label{tab:ai_k8s_comparison}
\resizebox{\textwidth}{!}{%
\begin{tabular}{|l|l|l|l|}
\hline
\textbf{Aspect} & \textbf{Kueue} & \textbf{DAS} & \textbf{llm-d and GAIE} \\
\hline
Purpose & Job queuing and batching & GPU/accelerator slicing & Distributed LLM inference \\
\hline
Resource Mgmt & Workload admission based on quotas & Dynamic GPU slices per pod & Pools of model servers \\
\hline
Scheduler Link & Uses admission hooks & Enforces slices via scheduler & Routes requests to optimal inference endpoints \\
\hline
Workloads & Batch/AI jobs needing fairness & GPU-as-a-Service + Data Science & LLM prefill/decode tasks \\
\hline
Isolation & Fair queuing, priorities & Strong GPU isolation & Through inference pools \\
\hline
Perf. Focus & Throughput + fairness & GPU utilization, low waste & Efficient parallel inference \\
\hline
Complexity & Light, native integration & Needs custom device plugins & Requires distributed inference layer \\
\hline
Extensibility & Generic; policy-extensible & AI-specific; NVIDIA MIG focus & Integrates with Kubernetes Gateway API \\
\hline
\end{tabular}
}
\end{table*}

Within each \textit{ClusterQueue}, \textit{Kueue} introduces the concept of a \textit{ResourceFlavor} to represent distinct types or configurations of underlying resources. Each \textit{ResourceFlavor} defines a specific capacity pool and \textit{ClusterQueues} can advertise multiple flavors simultaneously. When jobs are enqueued through a \textit{LocalQueue}, \textit{Kueue} maps them to the appropriate \textit{LocalQueue} and selects a \textit{ResourceFlavor} that satisfies the job’s requirements, effectively bridging namespace-level workloads with cluster-level resource scheduling.

\begin{figure}[htb!]
    \centering
    \ifarxiv
      \includegraphics[width=0.75\linewidth]{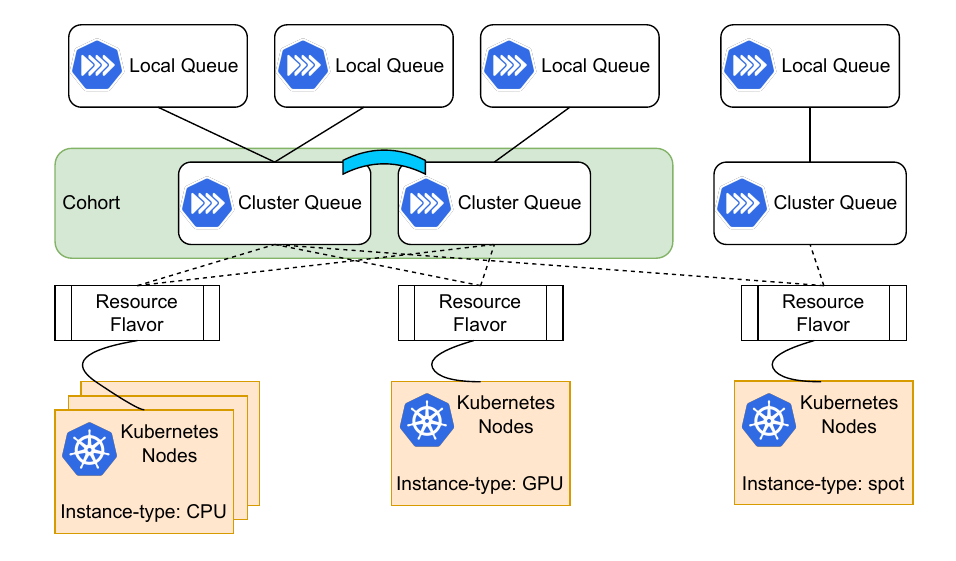}
    \else
    \includegraphics[width=\linewidth]{figures/kueue.pdf}
    \fi    
    \caption{Kueue working mechanism.}
    \label{fig:kueue}
\end{figure}

This hierarchical mapping (\emph{LocalQueue}~$\rightarrow$~\emph{ClusterQueue}~$\rightarrow$~\emph{ResourceFlavor}) forms the foundation of \textit{Kueue}’s scheduling architecture, enabling scalable, fair, and efficient sharing of heterogeneous compute resources. Figure~\ref{fig:kueue} illustrates the key \textit{Kueue} components and their interactions.

\textit{Kueue} addresses three key challenges: job admission delay, resource fragmentation, and fairness in multi-tenant scheduling. Its admission controller first evaluates whether a job can be admitted into its associated queue given current resource usage and configured limits. Once admitted, the job waits until scheduling slots become available in the \textit{ClusterQueue}.

\subsubsection{Accelerator slicing}

In Kubernetes-based AI inference deployments, a central challenge lies in efficiently partitioning and sharing GPU resources across workloads while maintaining performance isolation. To address this, OpenShift introduces the \textit{Dynamic Accelerator Slicer (DAS)}~\cite{das}, which provides on-demand, just-in-time GPU slice allocation driven by actual workload requirements.

Conventional GPU partitioning methods, such as static pre-slicing, often lead to resource fragmentation and suboptimal utilization. Moreover, static allocation performed at node initialization can cause service disruptions when workloads evolve and reconfiguration becomes necessary. In contrast, \textit{DAS} dynamically provisions and releases GPU slices in response to workload specifications, allocating resources only when a pod requests them and freeing them upon job completion. 

\textit{DAS} addresses four key challenges in achieving efficient, multi-tenant GPU utilization: it provisions slices just-in-time for new jobs, dynamically allocates and reclaims them as workloads change, coordinates with the NVIDIA GPU Operator to expose precise device configurations, and binds scheduling with slice allocation to ensure pods receive physical GPU slices only after successful provisioning. Together, these capabilities enable multiple inference workloads to share a single GPU efficiently, increasing utilization, reducing idle costs, and enhancing operational flexibility in shared AI-serving environments~\cite{singh2025}.

\subsubsection{Distributed Inference: llm-d and Gateway API Inference Extension}\label{sec:igw}

\textit{llm-d}~\cite{llmd} extends the Kubernetes ecosystem to efficiently orchestrate LLM inference workloads, addressing the limitations of conventional stateless load balancing. In Kubernetes, services paired with deployments typically distribute traffic evenly across identical replicas. This model is well-suited for microservices with short-lived, uniform requests, but LLM inference requests differ fundamentally: they are stateful, vary widely in compute cost, and benefit greatly from cache reuse and latency-aware routing \cite{llmd}.

To support these characteristics, the \textit{Gateway API Inference Extension (GAIE)} \cite{kubernetes-sigs-gateway-api-inference-extension} introduces first-class Kubernetes abstractions for managing distributed inference. \textit{GAIE} extends the standard Gateway API with specialized primitives that allow operators to declaratively expose, route and monitor model-serving endpoints. Its design bridges service networking and model serving, bringing inference-specific features such as model versioning, adaptive backend selection and prefix-cache awareness into the Kubernetes control plane.

\textit{GAIE} extends existing gateway concepts through two core components: (i) The \textit{InferenceGateway} acts as the primary entry point for inference traffic, augmenting a standard Gateway with capabilities for model registration, authentication, versioning, and routing policies—enabling multiple model versions to be served under a unified logical endpoint. (ii) The \textit{InferencePool} manages groups of backend model-serving replicas (e.g., \textit{vLLM}~\cite{kwon2023vllm}) and integrates an \textit{inference scheduler}, implemented in \textit{llm-d} as the \textit{Endpoint Picker (EPP)}, which intelligently routes each request to the most suitable backend, enabling cache-aware and latency-optimized inference scheduling within a Kubernetes-native framework.

A key efficiency driver in this design is the \textit{Key-Value (KV) cache}~\cite{llmd_kvcache}, an internal mechanism in transformer-based LLMs that stores intermediate representations (keys and values) for previously processed tokens. The initial \textit{prefill} phase encodes the full prompt into this cache, which is a computationally intensive step. Subsequent decoding stages reuse these cached representations, avoiding redundant computation and significantly reducing per-token latency. In distributed environments, if similar prompts are routed to different replicas without cache awareness, each must recompute the same prefix state, wasting GPU cycles and increasing latency. 

The \textit{llm-d} framework, through \textit{GAIE}’s programmable scheduling layer, optimizes routing based on cache locality and workload telemetry. By aligning inference request placement with active cache state, it reduces redundant computation, improves GPU utilization, and enhances throughput predictability.

\begin{table*}[htb]
\caption{Overview of the employed audio dataset, containing 16 files for medium model and 16 files for large model jobs from the Earnings 22 dataset benchmark~\cite{earnings22}. All values denote run time of the audio files to be transcribed.}
\label{tab:dataset}
\centering
\begin{tabular}{@{}lcccccc@{}}
\toprule
\textbf{Metric}              & \textbf{Count} & \textbf{Mean} [s] & \textbf{SD} [s] & \textbf{Median} [s] & \textbf{Min} [s] & \textbf{Max} [s] \\ \midrule
All Jobs & 32                        & 3127.38                 & 1217.90               & 3287.51                   & 1258.34                & 7405.44                \\
Large Model Jobs    & 16                        & 3077.63                 & 1584.01               & 3119.85                   & 1258.34                & 7405.44                \\
Medium Model Jobs   & 16                        & 3177.13                 & 742.37                & 3287.51                   & 1682.72                & 4380.08                \\ \bottomrule

\end{tabular}
\end{table*}

\subsubsection{Inference Runtime}

\textit{vLLM}~\cite{kwon2023vllm} is a high-performance and memory-efficient inference engine for LLMs. It is designed to address a key bottleneck in LLM serving: the inefficient use of GPU memory due to static and coarse-grained allocation in existing systems. \textit{vLLM} introduces \emph{PagedAttention}, a memory management mechanism inspired by virtual memory paging in operating systems. \textit{PagedAttention} enables fine-grained sharing of key-value (KV) caches across multiple inference requests, allowing dynamic batching and efficient context reuse without redundant memory copies. This design substantially increases GPU utilization and throughput while preserving low latency and model accuracy. Evaluations showed that \textit{vLLM} can achieve up to 3--5$\times$ higher throughput than baseline inference engines such as Hugging Face Accelerate or DeepSpeed-Inference under comparable latency targets \cite{kwon2023vllm}.

\subsection{Testbed Description}

All experiments were conducted on Red Hat OpenShift Service on AWS~\cite{rosa} (ROSA), a fully managed service that runs Red Hat's OpenShift enterprise Kubernetes platform with Platform-as-a-Service capabilities on Amazon Web Services. The cluster configuration used the default node types and counts for the control plane and infrastructure: 3 control plane nodes on \textit{m5.2xlarge} instances and 3 infrastructure nodes on \textit{r5.xlarge} instances. For the experiments, we added a single GPU node utilizing a \textit{p4d.24xlarge} instance deployed in the region \textit{us-west2}, having 8 NVIDIA A100 GPUs. The Whisper transcription jobs deployed on the GPU node are built on the Red Hat Universal Base Image~10 (UBI~10).

The cluster ran OpenShift version 4.19.16, which is based on Kubernetes 1.32. GPU management was handled by the NVIDIA GPU Operator v25.3.4. The complete technology stack is described in Section~\ref{sec:stack} and depicted in Figure~\ref{fig:architecture}. All the necessary files and configuration for replicating the test setup on any environment can be found on our GitHub repo~\cite{repo}.

\subsection{Test Cases \& Workload}
\label{sec:tests}

\begin{figure*}[htb!]
    \centering
    \begin{subfigure}[t]{0.48\textwidth}
        \centering
        \includegraphics[width=0.65\linewidth]{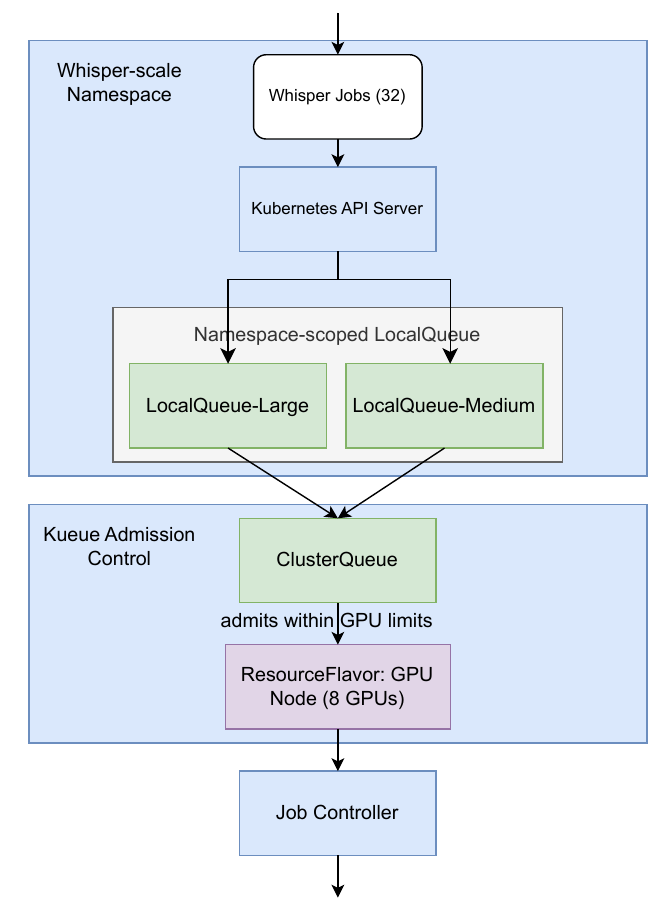}
        \caption{Transcription with Whisper with two local queues and one cluster queue.}
        \label{fig:kueue2}
    \end{subfigure}
    \hfill
    \begin{subfigure}[t]{0.48\textwidth}
        \centering
        \includegraphics[width=0.65\linewidth]{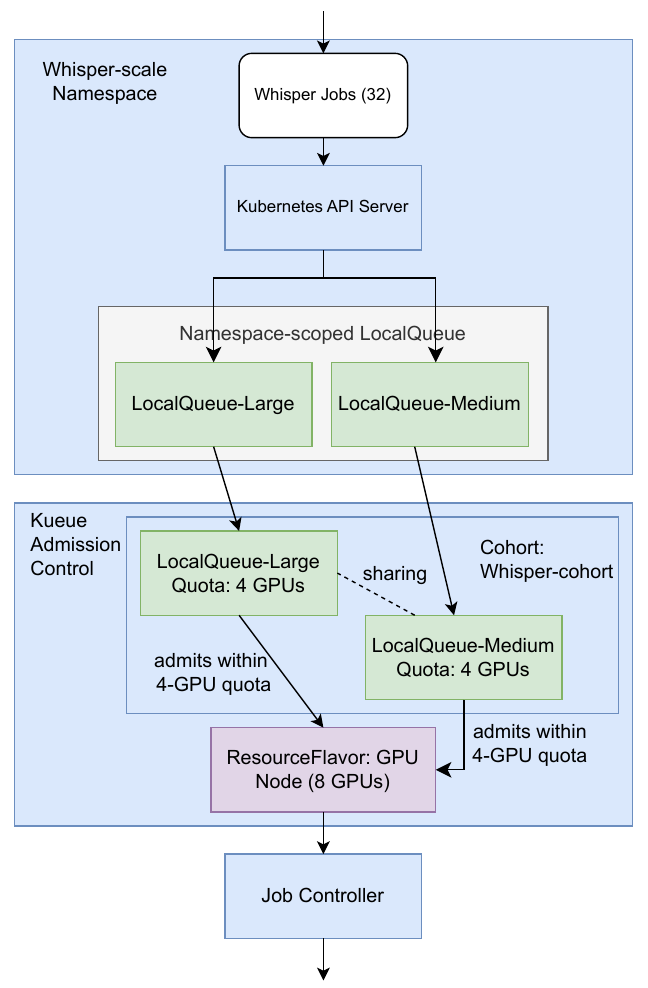}
        \caption{Transcription with Whisper with two local queues and two cluster queues and GPU borrowing.}
        \label{fig:kueue3}
    \end{subfigure}
    \caption{Kueue setup comparison for Whisper transcription experiments.}
    \label{fig:kueue_comparison}
\end{figure*}

To showcase how the Kubernetes and its ecosystem of projects can be combined to improve performance, we designed test cases for the multi-stage AI inference workflow. To run these test cases, we utilized the Earnings 22 dataset~\cite{earnings22}. This dataset comprises a corpus of 125 financial  earnings calls collected from various global companies. It serves as a commonly used benchmark for automatic speech recognition (ASR) models in real-world scenarios. We selected a subset of 32 files from this dataset: 16 files each to be transcribed by a job running the medium (769 M parameters) Whisper model and 16 files each to be transcribed by a job running the large (1550 M parameters) Whisper model. We selected two distinct models to mirror the industry trend of deploying models of varying sizes based on the use case, and to demonstrate how different MIG slice configurations can be utilized effectively with \textit{DAS}. For the same audio file, the large Whisper model is expected to take longer to transcribe than the medium model, since inference time scales approximately proportionally with model size and computational complexity. The dataset statistics are presented in Table~\ref{tab:dataset}.

\subsubsection{Scheduling}\label{sec:tests_kueue}
The first test case involves transcribing audio files from the dataset using \textit{Kueue} for job queue management. To achieve this, we had three different setups: (i) Not using \textit{Kueue} (to baseline with default Kubernetes job scheduling), (ii) using \textit{Kueue} with \textit{BestEffortFIFO} \textit{queueingStrategy} and 1 \textit{ClusterQueue} as outlined in Figure~\ref{fig:kueue2}, and (iii) using \textit{Kueue} with 2 cluster queues and resource borrowing as outlined in Figure~\ref{fig:kueue3}. The last setup allows us to borrow unused quota from the collective pool of the \textit{cohort} to accommodate more jobs. From a technical standpoint, each file transcription is deployed as a Kubernetes job that translates to a \textit {Kueue Workload} running a Whisper model, utilizing the entire GPU. 

\subsubsection{Accelerator Slicing}\label{sec:tests_das}
In the second test case, we employed \textit{DAS} to dynamically allocate GPU resources and execute transcription tasks using Kubernetes jobs that leverage Whisper models. Two configurations were evaluated: (i) without \textit{DAS}, where transcription jobs were distributed across eight full GPUs, and (ii) with \textit{DAS} enabled, where GPUs were partitioned into slices using the \textit{1g.5gb} instance profile for medium jobs and the \textit{3g.20gb} instance profile for large model jobs according to the Whisper model resource requirements~\cite{openai2022whispergithub}.

\subsubsection{Distributed Inference}\label{sec:tests_llmd}
\begin{figure}[htb!]
    \centering
    \ifarxiv
      \includegraphics[width=0.75\linewidth]{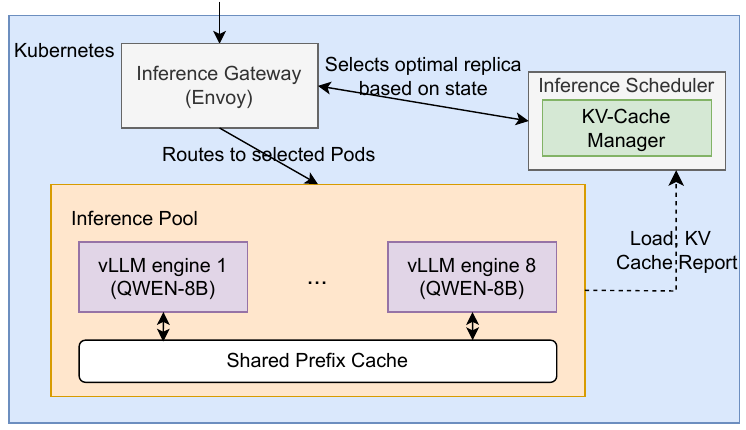}
    \else
    \includegraphics[width=\linewidth]{figures/llm-d.pdf}
    \fi            
    \caption{Overview of the llm-d setup.}
    \label{fig:llmd}
\end{figure}

In the final stage of our workflow, we evaluate distributed inference performance for summarization tasks using the open-weight and publicly released \textit{Qwen3-8B} model~\cite{qwen}, deployed on Kubernetes through the \textit{llm-d} stack. Specifically, we deploy eight replicas of \textit{vLLM}, each serving an identical Qwen3-8B model instance. The Qwen LLM family is developed by Alibaba, and we opted for this particular model due to its open-source nature and enterprise adoption. We first compare three configurations to quantify the overhead of \textit{GAIE}: (i) a standard \textit{Kubernetes ClusterIP} service for stateless load balancing, (ii) \textit{GAIE} with only the \textit{InferenceGateway}, and (iii) \textit{GAIE} with both the \textit{InferenceGateway} and \textit{InferencePool} enabled, including its \textit{inference-scheduler}. Additionally, we assess routing strategies within \textit{GAIE}, contrasting a random scheduler with the \textit{precise-cache ``well-lit path''}~\cite{llmd_welllit} strategy that leverages KV-cache telemetry to route requests toward replicas with matching prefix tokens. An overview of the setup is shown in Figure~\ref{fig:llmd}.


\subsection{Tooling}\label{sec:stack}

For test orchestration, we selected \textit{kube-burner}~\cite{10.1145/3629527.3651405} as our primary workload orchestration and analysis framework. Unlike generic load-testing utilities, kube-burner is purpose-built for Kubernetes and OpenShift environments, providing a Kubernetes-native approach to performance and scalability evaluation. Its core strength lies in a declarative, YAML-based configuration model that enables precise definition of complex object creation, modification, and deletion scenarios. This declarative design is critical for experimental rigor, offering powerful and flexible control over workload. It enables robust measurement capabilities, which elevate it from a simple stress tool to a comprehensive analysis framework. Its core strength is the built-in instrumentation to scrape Prometheus endpoints on the cluster before and after each test, capturing empirical data on system health and resource utilization. Beyond what is available in Prometheus, it includes several custom measurements like \textit{jobLatency}, which quantifies the total time from a job's creation to the completion of its pods, providing key performance indicators for our Kueue and DAS experiments.

For the final LLM powered transcript summarization test case, we used the open-source benchmarking framework \textit{GuideLLM}~\cite{guidellm2024} to orchestrate the execution of the test against our OpenAI-compatible Kubernetes hosted inference endpoint. By simulating real-world inference workloads, \textit{GuideLLM} enables users to assess the performance, resource requirements, and cost implications of deploying LLMs on various hardware configurations. We fed the transcript data from the previous stage to \textit{GuideLLM}, which used it as a custom dataset for generating prompts. The tool was sending these prompts at various concurrency levels (i.e., number of synchronous streams in parallel), allowing us to capture critical LLM inference performance metrics such as time-to-first-token (TTFT) and inter-token latency (ITL), and ensuring benchmark reproducibility.

\section{Evaluation}\label{sec:eva}

\begin{table*}[htb!]
\centering
\caption{Performance comparison of Kueue Configurations (96 jobs across 3 runs per configuration). All values are in minutes. Makespan median computed from 3 run-level makespans comprising 32 jobs per run; other medians computed from 96 jobs (32 jobs across 3 test runs) from each configuration.}
\label{tab:kueue-performance}
\resizebox{\textwidth}{!}{%
\begin{tabular}{@{}lcccccccc@{}}
\toprule
\multirow{2}{*}{\textbf{Configuration}} &
\textbf{Makespan} &
\textbf{Queue Time} &
\textbf{Exec Time} &
\multicolumn{2}{c}{\textbf{Medium Jobs}} &
\multicolumn{2}{c}{\textbf{Large Jobs}} \\
\cmidrule(lr){5-6} \cmidrule(lr){7-8}
 & \textbf{(Med, (Min--Max))} & \textbf{(Med / P95)} & \textbf{(Med)} & \textbf{Queue (Med)} & \textbf{Exec (Med)} & \textbf{Queue (Med)} & \textbf{Exec (Med)} \\
\midrule
Pure Jobs (Baseline)                  & 60.5 (59.1--60.5) & 0.0 / 0.0   & 28.4 & 0.0  & 23.6 & 0.0  & 35.3 \\
Kueue BestEffortFIFO                 & 62.5 (62.4--65.8) & 16.7 / 39.5 & 9.4  & 15.9 & 6.5  & 17.3 & 17.8 \\
+ Priority                     & 54.7 (52.2--56.1) & 19.5 / 44.5 & 9.5  & 39.1 & 6.7  & 10.2 & 17.1 \\
+ Preemption                   & 51.1 (49.2--52.7) & 25.0 / 41.9 & 8.9  & 36.7 & 6.4  & 3.7  & 15.6 \\
2 ClusterQueues + Borrowing & 55.0 (54.2--60.0) & 13.7 / 32.4 & 8.7  & 10.2 & 6.7  & 25.1 & 18.0 \\
\bottomrule
\end{tabular}
}
\end{table*}

To validate the proposed multi-stage inference use case, we conducted a series of experiments assessing the contribution of each component to the overall Generative AI (GenAI) workflow. The evaluation is structured to isolate and measure the performance gains delivered by \textit{Kueue} in batch scheduling, \textit{Dynamic Accelerator Slicer (DAS)} in GPU resource slicing and utilization, and the \textit{Gateway API Inference Extension (GAIE)} in low-latency distributed inference serving. Together, these results comprehensively illustrate how orchestrating Kubernetes-native primitives can efficiently support both large-scale and latency-sensitive AI workloads.

\subsection{Scheduling: Kueue}
\label{sec:scheduling-kueue}
To evaluate the scheduling and queueing behavior of Whisper transcription workloads under \textit{Kueue}, we conducted experiments as described in Section~\ref{sec:tests_kueue} both with and without Kueue enabled. We further explored two distinct Kueue configurations: (i) two \textit{LocalQueues} feeding into a single \textit{ClusterQueue}, and (ii) two \textit{LocalQueues} each mapped to separate \textit{ClusterQueues}. In all configurations, the cluster contained a total of eight GPUs, all corresponding to the same \textit{ResourceFlavor}. The key distinction between the two setups lies in resource sharing: in the first configuration, all eight GPUs are managed by a single \textit{ClusterQueue}, whereas in the second, each \textit{ClusterQueue} controls four GPUs, with one allowed to borrow resources from the other when idle capacity is available.  In this experimental setup, it is important to clarify the usage of the terms \textit{medium} and \textit{large}. These terms refer exclusively to the Whisper model variants used for transcription and do not denote the duration or size of the input audio files. Specifically, jobs labeled as \textit{large} use the Whisper \textit{large} model, which offers higher transcription accuracy at the cost of increased computational complexity and processing time. Consequently, job runtime is influenced by both the selected model variant and the input audio length; for example, a transcription using a \textit{medium} model on a longer audio file may take longer to complete than a \textit{large} model applied to a shorter audio clip.

For the first configuration, we evaluated three scheduling policies: (a) no priorities and no preemption, (b) higher priority assigned to large jobs through a \textit{WorkloadPriorityClass} but no preemption, and (c) higher priority assigned to large jobs, with preemption enabled for higher-priority workloads. In total, five distinct experimental setups were executed, each repeated three times. We used \textit{kube-burner} to create one large Whisper job followed by one medium Whisper job until 16 jobs of each kind were created for a total of 32 \textit{Kueue Workloads}. A key performance metric in our evaluation is the \textit{makespan}, which represents the total elapsed time required for all 32 transcription jobs to complete. The aggregated results across all runs are presented in Table~\ref{tab:kueue-performance}.

Our experimental results reveal a nuanced trade-off between performance optimization and scheduling predictability when comparing \textit{Kueue}-managed workloads to native Kubernetes job scheduling. While the pure-jobs scenario achieved a lower \textit{makespan} (60.5 min) compared to \textit{Kueue's BestEffortFIFO} (62.5 min), this improvement came at the cost of non-deterministic execution ordering. In both scenarios, the initial job execution was identical: the first eight jobs (\textit{medium-0} through \textit{medium-3} and \textit{large-0} through \textit{large-3}) initially consumed all available GPU resources. However, the subsequent scheduling of the 24 pending pods from the 24 pending jobs diverged significantly between the two approaches. This difference in execution order, rather than resource utilization efficiency, accounts for the observed variation in makespan.

\textit{Kueue}'s strict \textit{BestEffortFIFO} ordering produced predictable and reproducible scheduling patterns. Once the initial eight jobs occupied all GPU resources, \textit{Kueue} consistently scheduled jobs in the sequence: \textit{medium-4}, \textit{large-4}, \textit{medium-5}, \textit{large-5}, and so forth through \textit{medium-15} and \textit{large-15}. This deterministic behavior stems from \textit{Kueue}'s queue management logic, which maintains a stable ordering independent of the lower-level Kubernetes scheduler dynamics. In contrast, the pure-jobs scenario exhibited variable scheduling patterns influenced by the native Kubernetes scheduler's internal mechanisms, including its unschedulable and backoff queue policies. While the first eight jobs remained consistent, subsequent job ordering varied across runs---sometimes following the \textit{medium-4}, \textit{large-4} pattern, but other times deviating based on transient cluster state and scheduler heuristics.

This comparison underscores a fundamental tension in batch workload management: the native Kubernetes scheduler's opportunistic approach may occasionally yield a shorter \textit{makespan} through fortuitous ordering, but \textit{Kueue} provides guarantees of fairness, predictability, and reproducibility, which are qualities essential for multi-tenant and production environments. The deterministic behavior offered by \textit{Kueue} ensures that workload execution patterns remain consistent across runs, enabling reliable capacity planning, accurate service level agreement commitments, and equitable resource sharing among competing workloads. 

While the baseline scenario achieved a lower makespan than \textit{Kueue's  BestEffortFIFO}, there is a clear progression in scheduling efficiency and workload balance across successive Kueue configurations. Adding higher workload priority to the large Whisper jobs through \textit{WorkloadPriorityClass} reduced the makespan to 54.7 min and improved throughput for large jobs, reducing their queue times by nearly half, at the expense of longer delays for medium jobs. Enabling preemption along with higher workload priority provided the best improvement in overall makespan among all \textit{Kueue}-based configurations (51.1 minutes; an improvement of about 15\% compared to the baseline), as higher-priority jobs could interrupt and reclaim GPU resources, achieving faster turnaround without compromising execution efficiency. Interestingly, while large jobs gained most from this strategy, their longer execution durations still dominated the tail of the makespan distribution. It is important to note that while queueing time is effectively 0, median job execution times are higher (28.4 minutes) in the pure-jobs configuration because pods are created immediately after job submission, allowing all workloads to begin execution concurrently even when several pods are unschedulable due to lack of resources. In contrast, with \textit{Kueue}-managed scheduling, a workload is only admitted when adequate resources become available, after which the corresponding job and pod are created, leading to lower execution times (8.7--9.5 minutes). 

The final configuration, employing two \textit{ClusterQueues} with borrowing enabled, struck the best balance between fairness and utilization. It allowed underutilized GPUs from the medium \textit{ClusterQueue} queue to be dynamically reallocated once all medium jobs completed execution, improving medium job latency while maintaining near-optimal makespan. Together, these results demonstrate how combining queue management, priority, preemption, and resource sharing enables \textit{Kueue} to reduce total makespan while gracefully handling large, long-running jobs in heterogeneous AI workloads.

\textit{Kueue} introduces an admission control mechanism that determines when a workload can be accepted into execution based on the availability of cluster resources and the queue policy. To this end, we investigate the admission attempt durations, that is, the time between a workload’s admission attempt and \textit{Kueue}’s final admission response. Table \ref{tab:kueue-admission} summarizes the median (P50), 95th, and 99th percentile admission attempt durations across multiple \textit{Kueue} configurations. All values remain below 25 ms even at the 99th percentile, demonstrating that the queueing logic imposes negligible overhead on scheduling latency. Priority and preemption have little effect on admission latency, maintaining sub-10 ms decision times across all percentiles. The configuration with two \textit{LocalQueues} and two \textit{ClusterQueues} exhibits slightly higher tail latencies (P95 \textasciitilde 17 ms, P99 \textasciitilde 24 ms) due to cross-queue synchronization and resource borrowing coordination. Nonetheless, the absolute values remain operationally negligible compared to job execution times, showcasing that \textit{Kueue}’s admission control scales efficiently and can enforce advanced policies such as priority and borrowing without incurring measurable delay in workload admission.

\begin{table}[htb!]
\centering
\caption{Mean admission attempt duration across Kueue configurations (averaged over 3 runs). All values in milliseconds.}
\label{tab:kueue-admission}

\ifarxiv
\else    
\resizebox{\columnwidth}{!}{%
\fi
\begin{tabular}{@{}lC{1.3cm}C{1.3cm}C{1.3cm}@{}}
\toprule
\multirow{2}{*}{\textbf{Configuration}} &
\multicolumn{3}{c}{\textbf{Admission Attempt Duration (ms)}} \\
\cmidrule(lr){2-4}
 & \textbf{P50} & \textbf{P95} & \textbf{P99} \\
\midrule
Kueue BestEffortFIFO            & 2.67 & 7.64 & 20.9 \\
+ Priority                      & 2.51 & 4.77 & 6.48 \\
+ Priority + Preemption         & 2.57 & 5.48 & 12.8 \\
2 LocalQueues + 2 ClusterQueues & 2.82 & 17.0 & 23.7 \\
\bottomrule
\end{tabular}
\ifarxiv
\else    
}
\fi
\end{table}

\subsection{Accelerator Slicing: DAS}

To assess the impact of utilizing \textit{DAS}, we executed the second test case outlined in Section~\ref{sec:tests_das} in two different setups, one using \textit{DAS} to partition GPUs and another without it, having each job consume an entire GPU. In short, this test case aimed to quantify the efficiency gains obtained when using dynamic MIG-slicing to split the GPUs into smaller slices in order to run more than one Whisper transcription job concurrently. Theoretically, this method offers efficiency gains as each job no longer consumes an entire physical GPU. The GPU instance profiles have been selected depending on the Whisper model resource requirements~\cite{openai2022whispergithub}: for the medium model jobs the slice size had been set to \textit{1g.5gb} and for large model jobs to \textit{3g.20gb}. Similar to the Kueue experiments discussed in \ref{sec:scheduling-kueue}, the terms \textit{large} and \textit{medium} refer exclusively to the Whisper model variants used for transcription and have no relation to the duration or size of the input audio files.

At the time of this writing, the integration between the \textit{DAS} operator and Kueue for GPU memory aware scheduling of AI workloads was still in progress~\cite{instaslicepr910}, so the following experiments were orchestrated directly through Kubernetes native jobs utilizing \textit{DAS}. By exposing GPU memory as an extended resource, Kueue will be able to enforce quotas and admission control while \textit{DAS} continues to manage MIG slice allocation. The integration will focus on long-running workloads (Pods, Deployments, StatefulSets) commonly used for inference and serving applications.

Before investigating efficiency gains, we wanted to gain a thorough understanding of the impact of running a Whisper transcription job on a MIG slice as opposed to a full GPU. To this end, we conducted an experiment with only eight total jobs, half running the medium model and half running the large model. This setup ensured that waiting time was excluded from the job completion time measurement as both variants provided enough concurrency primitives: eight GPUs for no MIG setup and eight GPU slices for the MIG setup. Each configuration (no MIG vs. MIG) was run three times, the aggregated results are presented in Table~\ref{tab:ev:das_short}.

\begin{table}[htb!]
\caption{Influence of MIG slicing on transcription job performance with limited concurrency (24 jobs across 3 runs, each with 8 parallel executions).}
\label{tab:ev:das_short}
\centering
\begin{tabular}{@{}lll@{}}
\toprule
\textbf{Metric}                 & \textbf{No MIG} & \textbf{MIG}  \\ \midrule
Concurrent Jobs                 & 8    & 8   \\
Mean Job Completion Time, min          & 14   & 17 \\
Median Job Completion Time, min        & 10   & 13  \\
P95 Job Completion Time, min           & 36   & 40 \\
Std Dev, min                    & 11   & 11  \\
Mean GPU Utilization, \%         & 45   & 19   \\
Mean Tensor Core Utilization, \% & 1    & 1    \\
Mean DRAM Activity, \%           & 13   & 12   \\
Mean Worker CPU Utilization, \%  & 9    & 9 \\ \bottomrule
\end{tabular}
\end{table}

The mean job completion time increased from 14 minutes without slicing to 17 minutes with slicing, a similar increase is evident in the median (from 10 to 13 minutes) and the 95th percentile (from 36 to 40 minutes). The average GPU utilization dropped from 45\% without slicing to 19\% with slicing and all other relevant GPU metrics showed almost no difference. Running Whisper transcription jobs on proposed slice sizes introduced moderate but noticeable performance degradation due to resource constraints.

After understanding the performance characteristics of Whisper jobs on MIG slices, we conducted a second experiment with the complete dataset of 32 jobs to assess the potential gains that can be achieved with parallel execution utilizing \textit{DAS}. Again, each configuration (no DAS vs. DAS) was run three times and \textit{kube-burner} was used to create one medium Whisper model job followed by a large Whisper model job until 16 jobs of each kind were created for each run. The aggregated results are presented in Table~\ref{tab:ev:das}.

\begin{table}[htb!]
\caption{Influence of DAS on transcription job performance on the full dataset (96 jobs across 3 runs).}
\label{tab:ev:das}
\centering
\begin{tabular}{@{}lll@{}}
\toprule
\textbf{Metric}                 & \textbf{No DAS} & \textbf{DAS}  \\ \midrule
Concurrent Jobs                  & 8      & 25   \\
Mean Job Completion Time, min           & 28   & 18 \\
Median Job Completion Time, min         & 28   & 16  \\
P95 Job Completion Time, min            & 51   & 36 \\
Std Dev, min                    & 14    & 9  \\
Mean GPU Utilization, \%         & 44     & 38   \\
Mean Tensor Core Utilization, \% & 1      & 3    \\
Mean DRAM Activity, \%           & 13     & 38   \\
Mean Worker CPU Utilization, \%  & 10    & 37 \\ \bottomrule
\end{tabular}
\end{table}

\textit{DAS} dynamically increased the number of concurrently running transcription jobs from 8 (number of GPUs) to 25 (number of slices for the given specifications). This enhanced parallelism reduced the overall completion time of jobs, which is here equivalent to the sum of waiting time and execution time. The mean job completion time decreased from 28 minutes without DAS to 18 minutes with \textit{DAS}. Similar improvements are evident in the median (from 28 to 16 minutes) and the 95th percentile (from 51 to 36 minutes). Comparing these results with the first experiment, the execution parallelization achieved with \textit{DAS} and the resulting waiting times reduction essentially over-weighted the execution time degradation caused by resource constraints on MIG slices.

Increased execution parallelism also resulted in increased resource utilization. With \textit{DAS}, both the average tensor core utilization and the average DRAM activity were almost three times higher. Note that the moderate decrease of the average GPU utilization is due to the nature of MIG-slicing: with certain slice allocation patterns (e.g., two 3g.20gb GPU instance profiles scheduled on a single A100 GPU), the residual capacity may not be allocable, resulting in some streaming multiprocessors remaining idle and consequently lowering the overall utilization average. Despite this ``slice padding'', the overall idle capacity had been significantly reduced. Also the average CPU utilization on the worker node was nearly four times higher with \textit{DAS} (37\%) compared to the not using \textit{DAS} case (10\%).

\subsection{Distributed Inference: llm-d and Gateway API Inference Extension}

To evaluate distributed inference performance using \textit{llm-d} and the \textit{InferenceGateway} (via the Gateway API Inference Extension, GAIE), we adopted intelligent inference scheduling through the \textit{Precise Prefix cache aware} (also known as Precise Scheduling) ``well-lit path"" in the final stage of our workflow. The setup deployed eight replicas of vLLM, each serving the Qwen/Qwen3-8B model, behind an OpenShift Service Mesh \textit{envoy proxy}~\cite{envoyproxy} acting as the \textit{GAIE} gateway.

\begin{figure}[htb!]
    \centering
    \ifarxiv
      \includegraphics[width=0.75\linewidth]{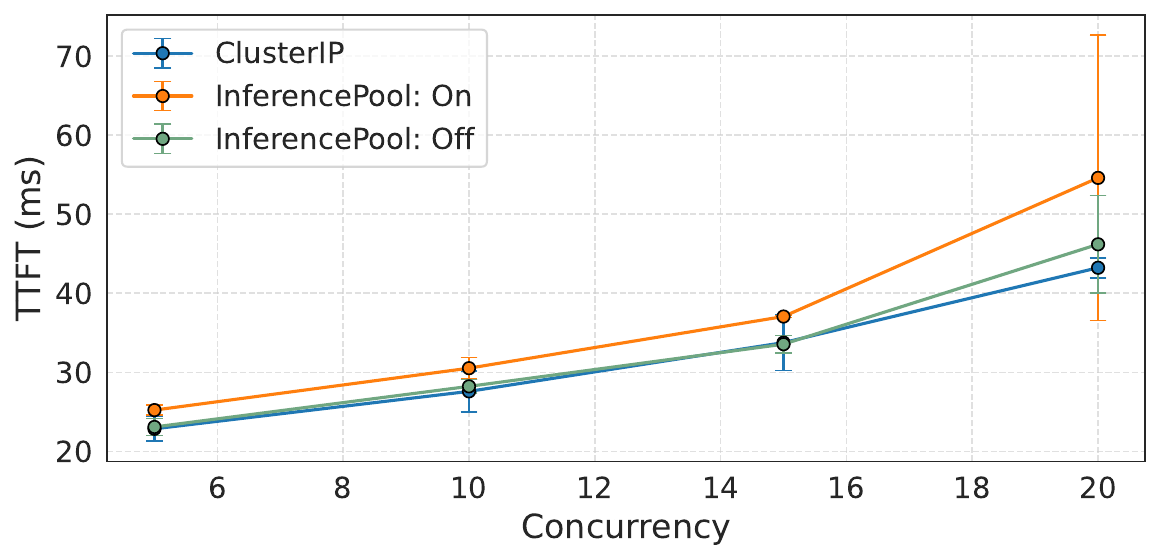}
    \else
    \includegraphics[width=\linewidth]{figures/overhead.pdf}
    \fi         
    \caption{Comparison of the overhead introduced by different scenarios. Error bars indicate 95\% confidence interval. }
    \label{fig:overhead}
\end{figure}

The first experiment investigates the latency overhead introduced by the \textit{GAIE} components as explained in Section~\ref{sec:tests_llmd}. To avoid potential latency introduced by the LLM runtime and accurately measure overhead, a single replica of the \textit{vLLM} Simulator~\cite{llmd_inference_sim} pod was used instead of a real \textit{vLLM} instance. Figure~\ref{fig:overhead} depicts the overhead where the x-axis is reflecting the concurrency level (the number of synchronous streams in parallel) and the y-axis the average Time to First Token (TTFT) in ms. Throughout the experiment, the \textit{ClusterIP} and \textit{Inference Pool}-disabled scenarios exhibit very similar TTFT. In contrast, in the \textit{Inference Pool}-enabled scenario this latency is between 3 ms and 11 ms higher compared to the other two setups. TTFT increases with concurrency and the \textit{Inference Pool}-enabled scenario exhibits the highest latency, especially at higher concurrency levels, introduced by additional routing overhead added by the gateway and endpoint picker components. This overhead is clearly offset by the latency improvements observed in the results that follow with real-world LLM use cases.

%
%

To evaluate the routing strategies within \textit{GAIE}, we compare two endpoint selection strategies: \textit{Random Scheduling} (Baseline) and \textit{Precise Scheduling} (\textit{``well-lit path''}). The workload comprised the 32 previously transcribed earnings call audio clips, each averaging 8{,}500 tokens (totaling 272{,}000 tokens). Using \textit{GuideLLM}, we generated 256 summarization requests (each of the 32 prompts repeated eight times) sent through the inference gateway, with a maximum output of 512 tokens per request for the summary. Each \textit{vLLM} replica ran on an NVIDIA A100 40GB GPU, with a per-instance KV-cache capacity of 143{,}360 tokens, yielding an aggregate of 1{,}146{,}880 tokens across all GPUs. As the total prompt volume exceeds the per-GPU cache capacity, random routing inevitably incurs cache misses and LLM prefilling overhead. In contrast, \textit{Precise Scheduling} maximizes the reuse of hot prefixes across replicas, demonstrating measurable gains in cache hit rate, end-to-end latency, and overall throughput.

\begin{figure}[htb!]
    \centering
    \ifarxiv
      \includegraphics[width=0.75\linewidth]{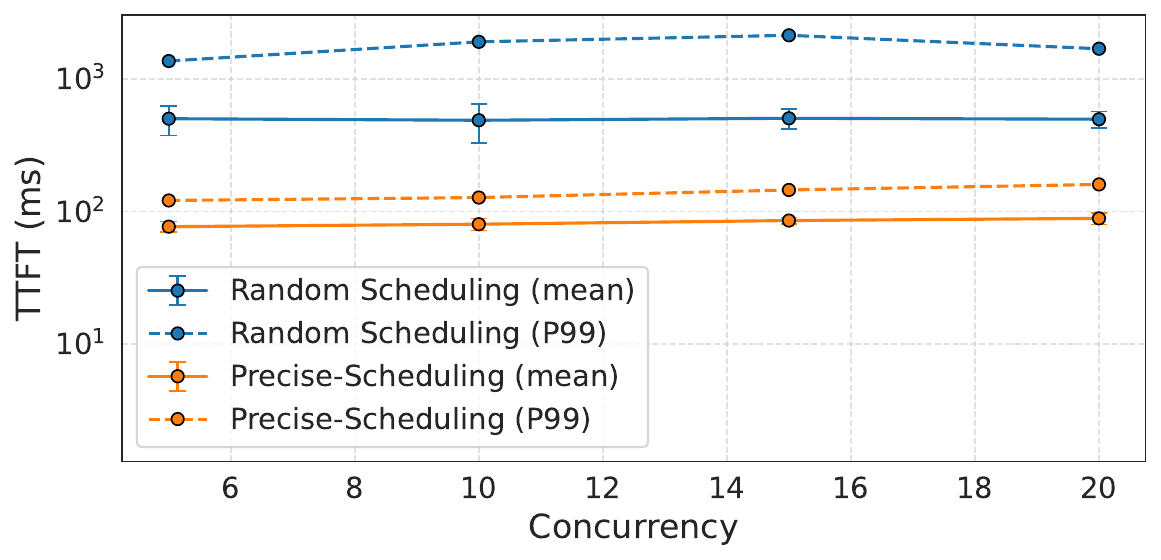}
    \else
    \includegraphics[width=\linewidth]{figures/ttft.pdf}
    \fi
    \caption{Comparison of the Time to First Token (TTFT). Error bars indicate 95\% confidence interval. Y-axis is depicted in log-scale. }
    \label{fig:ttft}
\end{figure}

Figure~\ref{fig:ttft} compares the TTFT between \textit{Random Scheduling} and \textit{Precise Scheduling}. The x-axis displays the synchronous concurrent streams in parallel, while the y-axis, in logarithmic scale, represents the TTFT in milliseconds. On average, \textit{Random Scheduling} ($\sim$500 ms) is approximately 6.25 times slower than \textit{Precise Scheduling} ($\sim$80 ms). Also, the variability is significantly higher for \textit{Random Scheduling}, with a 95\% confidence spread of 100 ms compared to about 6 ms for \textit{Precise Scheduling}. Additionally, the 99th percentile of TTFT for \textit{Random Scheduling} is up to 20 times slower than \textit{Precise Scheduling}.

\begin{figure}[htb!]
    \centering
    \ifarxiv
      \includegraphics[width=0.75\linewidth]{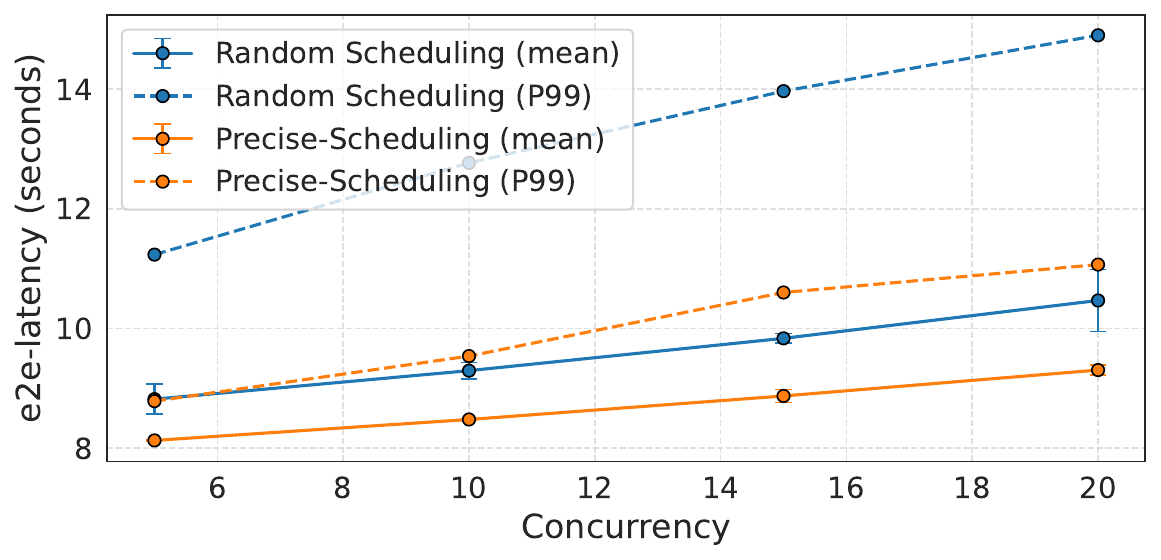}
    \else
    \includegraphics[width=\linewidth]{figures/e2e.pdf}
    \fi
    \caption{Comparison of the per request end-to-end request latency. Error bars indicate 95\% confidence interval.}
    \label{fig:e2e}
\end{figure}

Figure~\ref{fig:e2e} illustrates the per-request end-to-end request latency difference between \textit{Random Scheduling} and \textit{Precise Scheduling} routing strategies. The x-axis depicts the concurrency and the y-axis the latency. \textit{Random scheduling}, on average, incurs a latency of approximately 1 second more compared to \textit{Precise Scheduling}. Moreover, for the 99th percentile of end-to-end latency, \textit{Random Scheduling} can be up to 6 s slower. Additionally, it exhibits higher variability, as evident from the error bars and the significantly higher 99th percentile, suggesting that \textit{Precise Scheduling} provides better predictability.

\subsection{Takeaways}

Across all experiments, the results collectively demonstrate the complementary strengths of \textit{Kueue}, \textit{DAS}, and distributed inference scheduling (via \textit{GAIE}) in improving the efficiency, predictability, and scalability of AI inference workloads on Kubernetes. \textit{Kueue} provides deterministic, fair, and reproducible scheduling with negligible admission latency overhead, while priority and preemption mechanisms substantially reduce makespan, by up to 15\% and improve throughput for large jobs. \textit{DAS} enhances parallelism by dynamically partitioning GPUs into smaller slices, achieving up to a 36\% reduction in mean job completion time despite modest per-job slowdowns from resource sharing. Finally, \textit{GAIE}’s intelligent inference routing with \textit{Precise Scheduling} decreases latency, i.e., reducing 99th percentile TTFT by up to 90\% and end-to-end latency by up to 25\% , while ensuring stable, predictable performance even at the highest concurrency. Together, these findings highlight a clear design direction: combining queue-aware scheduling, dynamic accelerator partitioning, and intelligent inference routing yields an adaptive and efficient orchestration stack for large-scale, heterogeneous AI workloads, balancing throughput, fairness, and determinism across the stack.

\noindent

\section{Conclusion}\label{sec:conclusion}

This work demonstrates that Kubernetes, when extended with emerging AI-oriented projects, can effectively serve as a unified substrate for both batch and online Generative AI (GenAI) workloads. By combining Kueue, Dynamic Accelerator Slicer (DAS), and the Gateway API Inference Extension (GAIE), we have shown that the Kubernetes ecosystem is capable of orchestrating complex, multi-stage inference pipelines with improved utilization, scalability, and responsiveness. Our showcase, Whisper-based transcription and LLM-driven summarization, illustrates that container-native design principles can be successfully applied to heterogeneous, GPU-accelerated environments traditionally dominated by bespoke AI infrastructure.

Beyond technical integration, our evaluation reveals the operational synergies among these components: Kueue optimizes batch job queueing (up to 15\%), DAS maximizes accelerator efficiency through fine-grained GPU slicing (up to 36\%), and GAIE with the llm-d-inference-scheduler implementation delivers optimized Time to First Token (up to 82\%). Together, these elements represent a cohesive and forward-looking framework for self-hosting GenAI inference workloads in production-grade Kubernetes clusters. We conclude that the rapid evolution of Kubernetes-native AI extensions signals a maturing ecosystem---one that is increasingly capable of addressing the stringent performance, scalability, and manageability demands of modern AI systems.

\section{Future Work}\label{sec:future}

Future work will focus on extending the evaluation to encompass a broader set of AI workloads, including multimodal and streaming inference scenarios, to further validate the flexibility and scalability of the proposed workflow. As previously stated, native integration between Kueue and Dynamic Accelerator Slicing (DAS) is currently lacking; therefore we plan to have future work focusing on characterizing an end-to-end batch inference workflow that combines Kueue for workload queueing and DAS for accelerator slicing within Kubernetes. Additionally, we plan to integrate and assess emerging Kubernetes-native components for AI inference, such as gang scheduling and Dynamic Resource Allocation (DRA), to enhance the system’s efficiency and responsiveness under dynamic workloads. These efforts aim to generalize the presented framework and position it as a foundational blueprint for end-to-end AI workload orchestration on cloud-native infrastructure.

\clearpage 

\bibliographystyle{unsrt}  
\bibliography{ref}

\end{document}